\documentclass[12pt,a4paper,dvips]{article}
\usepackage{a4p}
\usepackage{cite,mcite}
\usepackage{graphicx}
\usepackage{physics }
\usepackage{l3_title,ifthen}
\journalname{Phys. Lett. B}
\date{May 5, 2000}
\preprint{-2000-061}
\newlength{\capindent}
\newlength{\capwidth}
\newlength{\figwidth}
\newcommand{\icaption}[2][!*!,!]{\hspace*{\capindent}%
  \begin{minipage}{\capwidth}
    \ifthenelse{\equal{#1}{!*!,!}}%
      {\caption{#2}}%
      {\caption[#1]{#2}}
  \end{minipage}}
\newcommand{\QQ}{\ensuremath{\mathrm{q} \bar{\mathrm{q}}}}

\newcommand{\DD}{\ensuremath{\mathrm{d} \bar{\mathrm{d}}}}

\newcommand{\UU}{\ensuremath{\mathrm{u} \bar{\mathrm{u}}}}

\newcommand{\EE}{\ensuremath{\mathrm{e}^+ \mathrm{e}^-}}

\newcommand{\FF}{\ensuremath{f\bar f}}

\newcommand{\EEFF}{\ensuremath{\EE\rightarrow\FF}}
\newcommand{\EEHA}{\ensuremath{\EE\rightarrow\mathrm{hadrons}}}

\newcommand{\COST}{\ensuremath{\cos \theta}}

\newcommand{\gL}{\ensuremath{g_\mathrm{L}}}
\newcommand{\gR}{\ensuremath{g_\mathrm{R}}}
\newcommand{\gLR}{\ensuremath{g}}

\newcommand{\LL}{\ensuremath{\mathrm{LL}}}
\newcommand{\RR}{\ensuremath{\mathrm{RR}}}
\newcommand{\RL}{\ensuremath{\mathrm{RL}}}
\newcommand{\LR}{\ensuremath{\mathrm{LR}}}

\newcommand{\LQ}{\ensuremath{\mathrm{LQ}}}

\def\SNM{\mathrm{\stackrel{\sim}{\nu}_{\mu}}}
\def\SNT{\mathrm{\stackrel{\sim}{\nu}_{\tau}}}

\newcommand{\ef}{\ensuremath{\mathrm{e}f}}
\begin{document}
\begin{titlepage}
\title{Search for Manifestations of New Physics in \\
        Fermion--Pair Production at LEP}
\author{The L3 Collaboration}
%
% The abstract
%
\vspace{-1.0cm}
\begin{abstract}
\noindent
The measurements of hadron and lepton--pair production cross sections and
leptonic forward--backward asymmetries performed with the L3 detector at
centre--of--mass energies between 130~{\GeV} and 189~{\GeV} are used to search
for new physics phenomena such as: 
contact interactions, 
exchange of virtual leptoquarks, scalar quarks and scalar neutrinos,
effects of {\TeV} strings in models of quantum gravity with large extra dimensions and
non--zero sizes of the fermions.
No evidence for these phenomena is found and
new limits on their parameters are set.
\end{abstract}
%
% Adds "To be submitted to ..." or "Submitted to ...", if relevant
%
\submitted
\end{titlepage}
%
%%%%%%%%%%%%%%%%%%%%%%%%%%%%%%%%%%%%%%%%%%%%%%%%%%%%%%%%%%%%%%%%%%%%%%%%%%%%%%%
% Introduction
%%%%%%%%%%%%%%%%%%%%%%%%%%%%%%%%%%%%%%%%%%%%%%%%%%%%%%%%%%%%%%%%%%%%%%%%%%%%%%%
%
\newpage
\section*{Introduction}

The study of fermion--pair production, {\EEFF},
at centre--of--mass energies 
well above the Z resonance 
allows to look for physics beyond the Standard Model.
The successful running of LEP in 1997 and 1998 at %higher
energies of 182.7~{\GeV} and 188.7~{\GeV}, and
the tenfold increase of luminosity compared to our previous 
searches~\cite{l3-131,l3-143}, improves substantially the sensitivity
to new physics phenomena.

The results presented in this paper are based on analyses of our
measurements of hadronic and leptonic cross sections and
leptonic forward--backward asymmetries~\cite{l3-90,l3-117,l3189pr}.
The measurements in all channels are used to search for 
%the existence of
four-fermion contact interactions. 
The virtual exchange of leptoquarks and
scalar quarks is  investigated using our hadron cross section measurements.
The effects of scalar neutrino exchange are looked for in all leptonic channels.
Limits on contact interactions, and on leptoquark, scalar quark and scalar neutrino
couplings have been presented also by other LEP
collaborations~\cite{aleph183,opal189}.
The effects of {\TeV} strings, predicted
recently~\cite{Antoniadis,Peskin} in theories of quantum gravity with extra
dimensions~\cite{ADD}, are searched for in Bhabha scattering.
This is an extension of the searches for low scale gravity in fermion--pair
production at LEP~\cite{opal189,l3gr1}.
Furthermore, a form factor ansatz is used to estimate the size
of leptons and quarks.

%%%%%%%%%%%%%%%%%%%%%%%%%%%%%%%%%%%%%%%%%%%%%%%%%%%%%%%%%%%%%%%%%%%%%%%%%%%%%%%%
%                             MEASUREMENTS
%%%%%%%%%%%%%%%%%%%%%%%%%%%%%%%%%%%%%%%%%%%%%%%%%%%%%%%%%%%%%%%%%%%%%%%%%%%%%%%%
\section*{Data and Analysis Method}

Measurements of cross sections and forward--backward asymmetries for the
reactions {\EEFF} have been performed with the 
L3 detector~\cite{l3-00}
at centre--of--mass
energies, $\sqrt{s}$, of 130.0~{\GeV}, 136.1~{\GeV}, 161.3~{\GeV}, 172.3~{\GeV},
182.7~{\GeV} and 188.7~{\GeV}~\cite{l3-90,l3-117,l3189pr}.
They correspond to an integrated luminosity of 265.4~pb$^{-1}$.

For the {\EE} final state both leptons have to be in the polar angular range
$44^{\circ}<\theta<136^{\circ}$, where $\theta$ is the angle between the
incoming electron and the outgoing lepton. Muon-- and tau--pair candidates are
selected with both leptons in the fiducial volume given by $|\COST|<0.9$
and $|\COST|<0.92$, respectively.
Hadron events are selected in the full solid angle.

In total 28470 hadron events and 9417 lepton--pair events are selected.
A minimum effective centre--of--mass energy, $\sqrt{s'_{\mathrm{min}}}$, 
or a maximum acollinearity angle in the Bhabha channel, are required to
select events without substantial energy loss due to initial state
radiation.
The remaining samples, which are studied in this paper,
contain in total 7785 hadron and 7704 lepton--pair events.

%%%%%%%%%%%%%%%%%%%%%%%%%%%%%%%%%%%%%%%%%%%%%%%%%%%%%%%%%%%%%%%%%%%%%%%%%%%%%%%%
%                             Analysis Method
%%%%%%%%%%%%%%%%%%%%%%%%%%%%%%%%%%%%%%%%%%%%%%%%%%%%%%%%%%%%%%%%%%%%%%%%%%%%%%%%
%\section*{Analysis Method}
The measurements of total cross sections
and leptonic forward--backward asymmetries are analysed in terms of new physics,
which will manifest itself as deviations from the Standard Model  predictions.
The contributions of contact interactions, leptoquarks and scalar quarks
are included directly into the improved Born
cross section calculated with the program ZFITTER~\cite{ZFITTERNPH} and are
convoluted to account for QED radiative corrections.
For the analyses including the {\EE} final state, i.e. contact interactions,
scalar neutrinos, {\TeV} strings and form factors,
the effects of new phenomena are computed with dedicated programs
in the improved Born approximation, taking into account QED radiative corrections.
For contact interactions where both approaches are used, the results
agree well with each other.

The measurements are compared to the predictions of the 
Standard Model~\cite{Standard_Model_ci} as calculated using the
ZFITTER and TOPAZ0~\cite{TOPAZ0} programs with the following
parameters~\cite{l3-69,PDG98,mt_tevatron}: 
$\MZ=91.190$~{\GeV},
$\alpha_{s}(\MZ^2)=0.119$,
$\Mt=173.8$~{\GeV},
$\Delta\alpha_{\mathrm{had}}^{(5)}=0.02804$,
and $\MH=150$~{\GeV}.
The results of our analyses are not sensitive to small variations of these 
parameters.
The theoretical uncertainties on the Standard Model
predictions are estimated to be below 1\% 
%~\cite{err-zf}
except for
large angle Bhabha scattering where the uncertainty is
2\%~\cite{TOPAZ0}.

The measurements show no statistically significant deviations from the
Standard Model expectations.
In their absence, limits at 95\%
confidence level on the contributions of new physics are determined
by integrating the log-likelihood functions
in the physically allowed range of the parameters describing
new physics phenomena, assuming a uniform prior distribution.
The statistical errors and systematic uncertainties of the
measurements~\cite{l3189pr,l3-117}, as well as the
theory uncertainties given above, are combined in quadrature
for all analyses.

%%%%%%%%%%%%%%%%%%%%%%%%%%%%%%%%%%%%%%%%%%%%%%%%%%%%%%%%%%%%%%%%%%%%%%%%%%%%%%%%
%                             RESULTS
%%%%%%%%%%%%%%%%%%%%%%%%%%%%%%%%%%%%%%%%%%%%%%%%%%%%%%%%%%%%%%%%%%%%%%%%%%%%%%%%
%===============================================================================
\section*{Four--Fermion Contact Interactions}
Four-fermion contact interactions offer a
general framework for describing interactions beyond the Standard Model. They
are characterised by a coupling strength, $g$, and by an energy scale,
$\Lambda$, which can be viewed as the typical mass of new heavy particles being
exchanged. At
energies much lower than $\Lambda$, the exchange of virtual new particles is
described by an effective Lagrangian~\cite{theory_ci}:

\begin{equation}
  \label{lagr_ci}
  {\cal L} = \frac{1}{1+\delta_{\ef}}\sum_{i,j = \mathrm{L,R}} \eta_{ij}
             \frac{g^2}{\Lambda^2_{ij}}
             (\bar{\mathrm{e}}_i \gamma^{\mu}\mathrm{e}_i)
             (\bar{\mathit{f}}_j \gamma_{\mu}\mathit{f}_j),
\end{equation}
where e$_i$ and $f_j$ denote the left-- and right--handed initial--state
electron and final--state fermion fields. The Kronecker symbol, 
$\delta_{\ef}$, is zero except for
the {\EE} final state where it is one. The parameters $\eta_{ij}$ define the
contact interaction model by choosing the helicity amplitudes which contribute
to the reaction $\EE\rightarrow\FF$.  The value of $g/\Lambda$ determines the
size of the expected effects. By convention $g^2/4\pi$ is chosen to be 1 and
$|\eta_{ij}|=1$ or $|\eta_{ij}|=0$, leaving the energy scale $\Lambda$ as a free
parameter. 
The helicity combinations of the specific models considered are
defined in Table~\ref{tab:ci-models}.
Atomic physics parity violation experiments probe with high precision the
couplings of electrons to quarks of the first family, and place severe
constraints on the scale $\Lambda$ of the order of
15~{\TeV}~\cite{atomic_parity_const}.  The VV, AA, V0 and A0 models are parity
conserving and hence are not constrained by such measurements.

\begin{figure}[htbp]
  \begin{center}
  \begin{tabular}{cc}
     \includegraphics[width=0.43\textwidth]{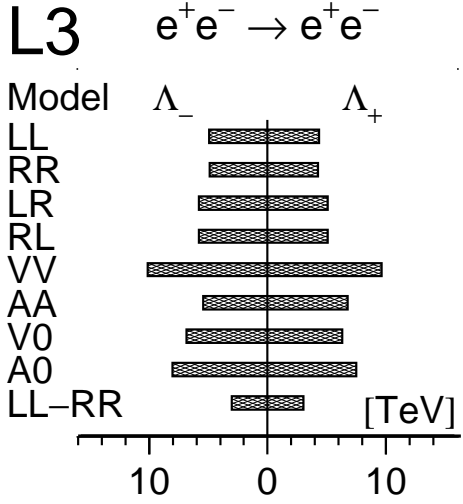} &
     \includegraphics[width=0.43\textwidth]{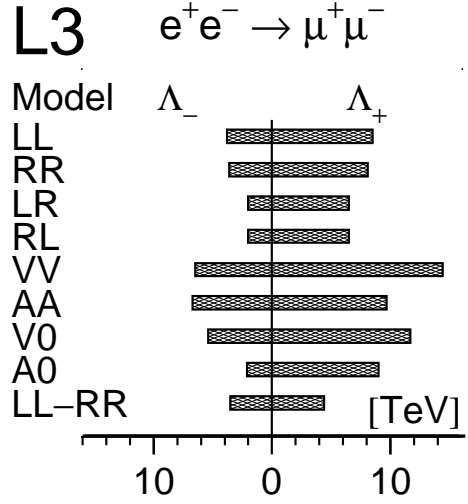} \\
     \includegraphics[width=0.43\textwidth]{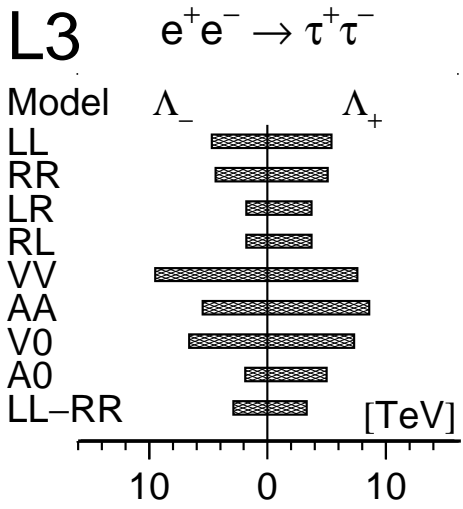} &
     \includegraphics[width=0.43\textwidth]{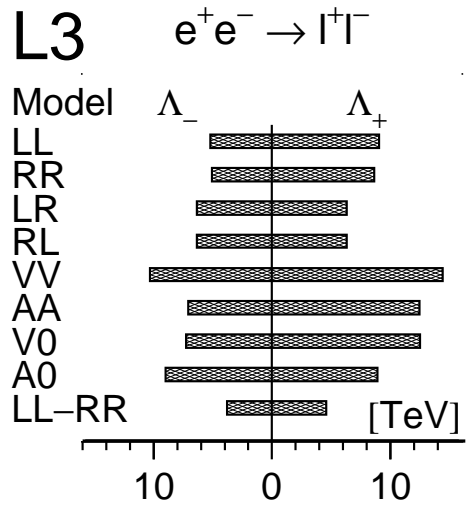} 
  \end{tabular}
  \end{center}
  \caption[]{
    One--sided 95\% confidence level lower limits on the scales $\Lambda_{+}$ 
    and $\Lambda_{-}$ for
    contact interactions in leptonic channels. The limits correspond to the
    values given in Table~\ref{tab:ci-leptons}.}
  \label{fig:ci-leptons}
\end{figure}

The four--fermion contact interactions for the different types of final--state
fermions are tested separately as well as for all flavours combined and lower
limits on the scale $\Lambda$ are derived. 
The lower limits on $\Lambda$ obtained from  
lepton--pair final states
are summarised in Table~\ref{tab:ci-leptons} 
and Figure~\ref{fig:ci-leptons}.
It is important to note that the pure leptonic case is only accessible at LEP.

\begin{figure}[htbp]
  \begin{center}
  \begin{tabular}{cc}
     \includegraphics[width=0.43\textwidth]{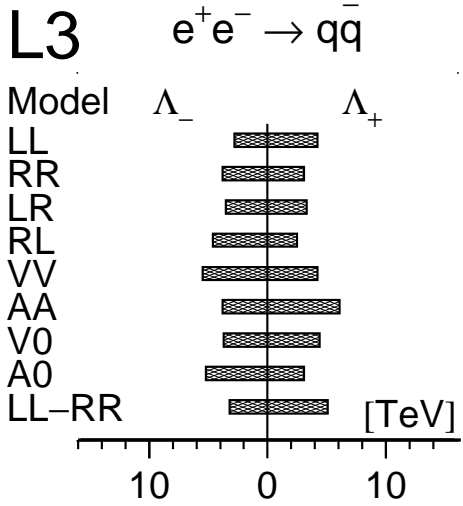} &
     \includegraphics[width=0.43\textwidth]{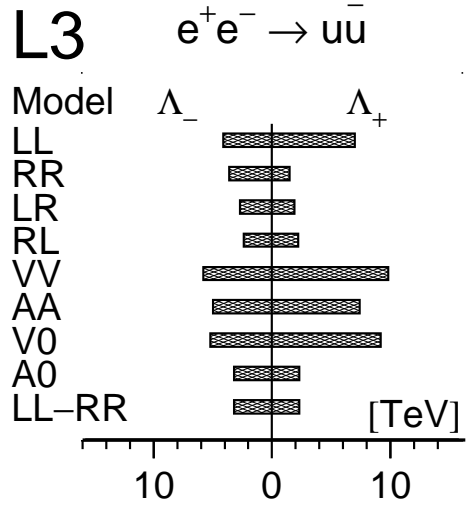} \\
     \includegraphics[width=0.43\textwidth]{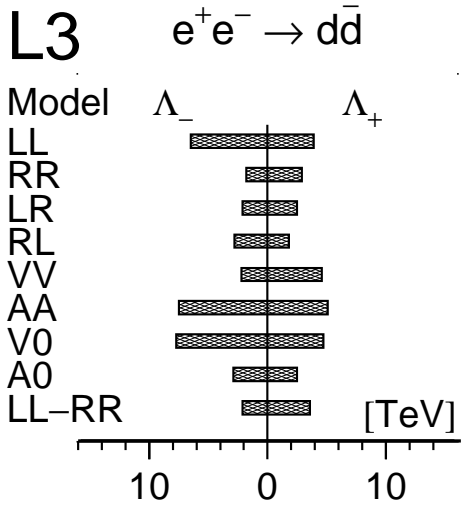} &
     \includegraphics[width=0.43\textwidth]{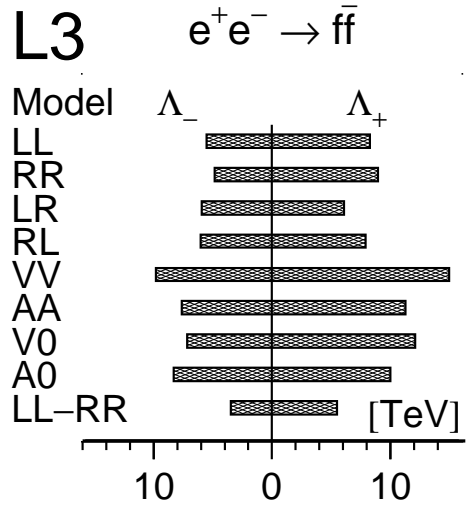} 
  \end{tabular}
  \end{center}
  \caption[]{
    One--sided 95\% confidence level lower limits on the scales $\Lambda_{+}$
    and $\Lambda_{-}$ for
    contact interactions in hadronic channels and in all channels combined.  The
    limits correspond to the values given in Table~\ref{tab:ci-hadrons}.}
  \label{fig:ci-hadrons}
\end{figure}

For hadronic final states the cases where the contact interactions
affect either all flavours at the same time,
or only one flavour of up--type or down--type quarks, are analyzed.
The results
are given in Table~\ref{tab:ci-hadrons} and depicted in
Figure~\ref{fig:ci-hadrons}, together with the
combined results for all charged fermions.
Similar limits are obtained from studies of
deep inelastic scattering at HERA~\cite{zeus-99,h1-00}
and proton--antiproton collisions at the TEVATRON~\cite{cdf-97,d0_ci}.

%===============================================================================
\section*{Leptoquarks}

Leptoquarks couple to quark--lepton pairs from the same family,  preserving the
baryon number $B$ and the lepton number $L$. Leptoquarks carry fermion numbers, 
$F=L+3B$. 
Following the notation in Reference~\cite{kalibib_9}, scalar leptoquarks $S_I$
and vector leptoquarks $V_I$ are indicated based on spin and isospin $I$.
Isomultiplets with different hypercharges are denoted by an additional tilde.
  
In the process {\EEHA}, leptoquarks of
the first generation can be exchanged in the $t$--channel ($F=0$) or in the 
$u$--channel ($F=2$).  
The coupling of leptoquarks to quark--lepton pairs, $g$, is referred to  as $\gL$ or $\gR$, 
according to the chirality of the lepton. 
The contributions of leptoquark exchange to  {\EE} $\rightarrow$ {\QQ} depend on $\gLR^2$.
 
\begin{figure}[htbp]
  \begin{center}
     \includegraphics[width=0.9\textwidth]{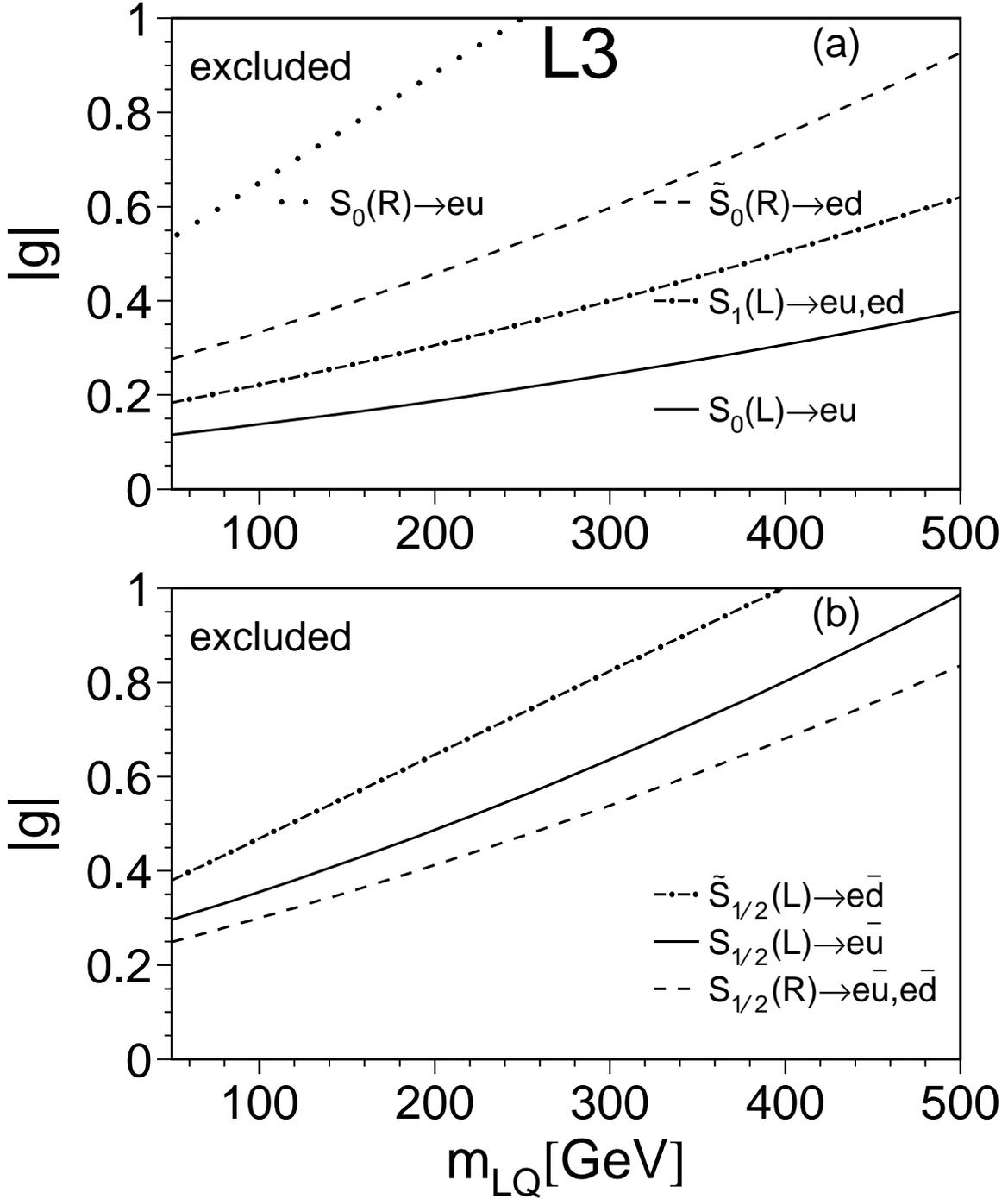}
  \end{center}
  \caption[]{
    The 95\% confidence level upper limits on $|g_L|$ or $|g_R|$ as a
    function of $m_{\LQ}$ for various scalar leptoquarks derived from hadronic
    final state cross sections.  
    Limits are shown for fermion number F=2 (a) and for F=0 (b).
    Bounds on the R--parity violating couplings $|\lambda'_{1jk}|$
    for the exchange of scalar down--type quarks in
    the $u$--channel and scalar up--type quarks in the $t$--channel correspond to
    limits on $|\gL|$ for the $S_0$(L) and $\tilde{S}_{1/2}$ leptoquark exchange,
    respectively.}
  \label{fig:sca}
\end{figure}

\begin{figure}[htbp]
  \begin{center}
     \includegraphics[width=0.9\textwidth]{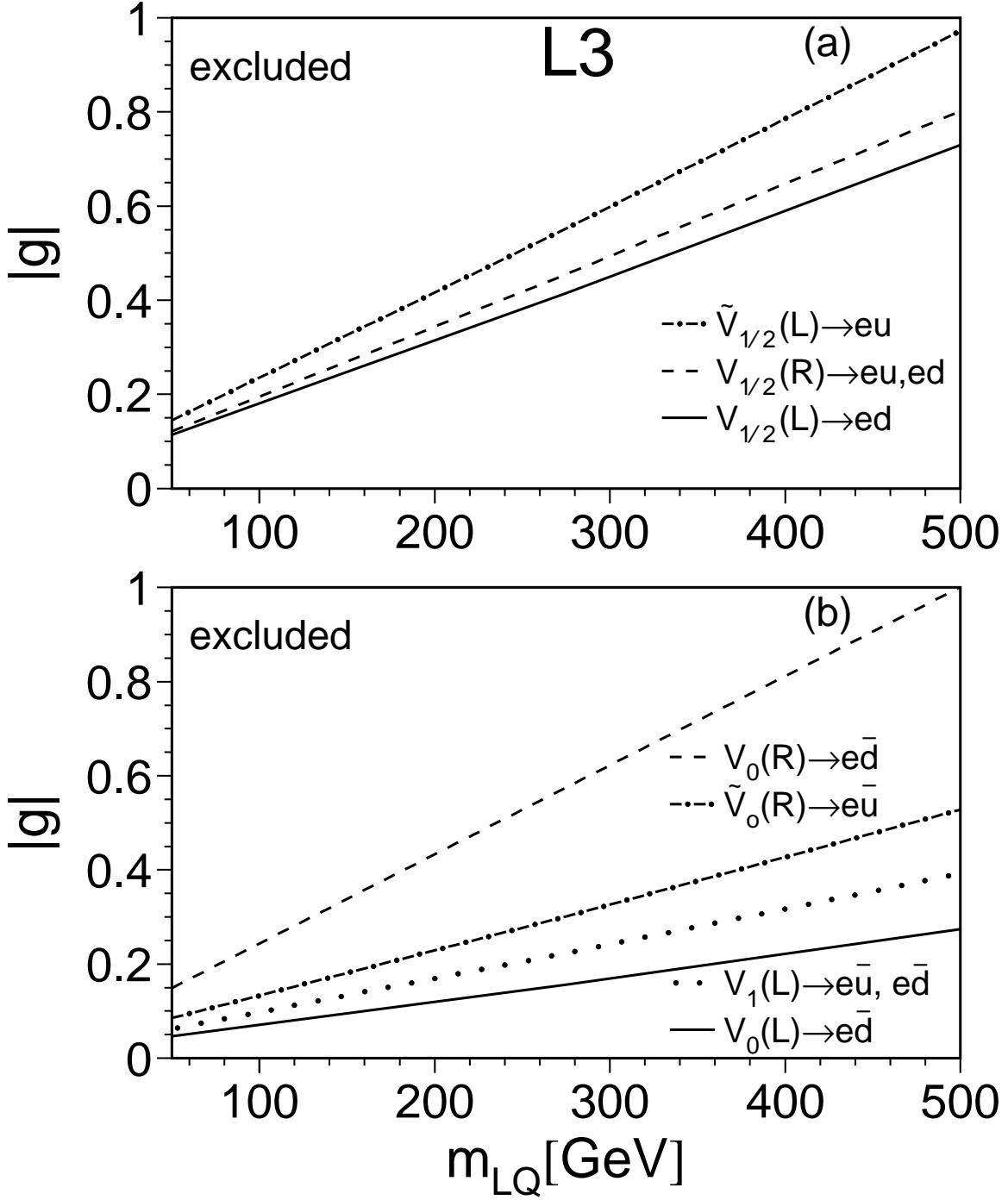}
  \end{center}
  \caption[]{
    The 95\% confidence level upper limits on $|g_L|$ or $|g_R|$ as a
    function of $m_{\LQ}$ for various vector leptoquarks derived from hadronic
    cross sections.  
    Limits are shown for fermion number F=2 (a) and for F=0 (b).}
   \label{fig:vec}
\end{figure}

Studying  the exchange of different types of leptoquarks separately,
limits on $|\gLR|$ are derived 
depending on the mass, $m_{\LQ}$, of the exchanged
leptoquark.  The states $S_{0}$, $S_{1/2}$ and 
$V_{0}$, $V_{1/2}$ couple to both left- and right-handed quarks.
Here, only $\gL$ or $\gR$ is assumed to be non--zero since low energy processes
and rare decays of $\pi$ and K constrain the product
$\gL\,\gR$~\cite{atomic_parity}.
Upper limits on the allowed values for $|\gLR|$ are presented in 
Figure~\ref{fig:sca} for scalar leptoquarks and in
Figure~\ref{fig:vec} for vector leptoquarks.
 
For a coupling of electromagnetic strength, 
\mbox{$\gLR = \sqrt{4\pi \alpha}$}, where $\alpha$ is the 
fine-structure constant, mass limits can be derived. The results for these lower 
bounds on leptoquark masses are given  in Table~\ref{tab:lqmass}.
In case of $\tilde{S}_{1/2}(\mathrm{L})$ exchange, i.e. coupling to left-handed fermions,
the assumption $\gL = \sqrt{4\pi \alpha}$ yields a very small
contribution to the hadron cross section that is not observable with the
precision of our measurements.

The results at LEP complement the leptoquark searches at HERA.
In most cases the indirect limits on leptoquark masses and couplings obtained
in our analysis are more stringent than the corresponding limits
presented by the H1 collaboration~\cite{h1-00}. 
The indirect search covers regions at high leptoquark masses
above the reach of direct leptoquark
searches~\cite{h1-lq-99}.

%===============================================================================
\section*{R--Parity Violating Scalar Neutrinos and Scalar Quarks}
Even in a minimal supersymmetric model~\cite{mysusy} the most general
superpotential contains
interactions violating R--parity in the trilinear couplings of
superfields. The only renormalisable gauge invariant operator that
couples fermions and their scalar partners is given
by~\cite{dimopoulos}:
\begin{equation}
 \label{eq:wsusy}
 W_{\not\! R} =
   \lambda _{ijk}       L_i        L_j   \bar{E}_k +
   \lambda'_{ijk}       L_i        Q_j   \bar{D}_k +
  \lambda''_{ijk} \bar{U}_i  \bar{D}_j   \bar{D}_k,
\end{equation}
where $L$ and $E$ are the leptonic,
and $Q$, $U$ and $D$ are the quark superfields.
The family indices are $i$, $j$ and $k$, e.g. $\lambda_{121}$ for $\SNM$
exchange in the reaction $\EE \rightarrow \EE$.

\begin{figure}[htbp]
  \begin{center}
     \includegraphics[width=0.999\textwidth]{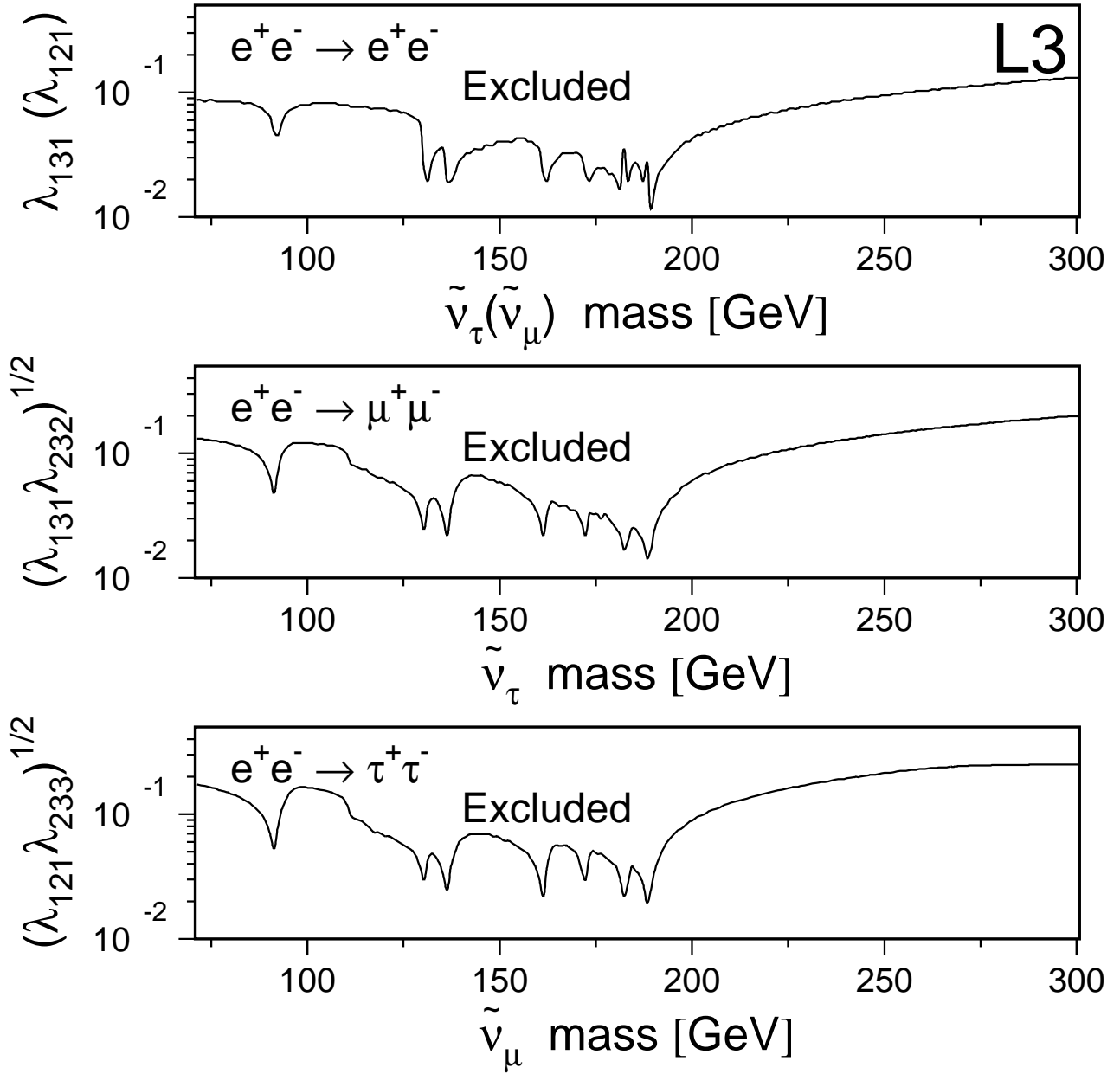}
  \end{center}
  \caption[]{
    Upper limits at 95\% confidence level on the coupling strengths
    $ \lambda_{ijk}$
    of scalar leptons to leptons as a function of the scalar
    neutrino mass, derived
    from measurements of lepton-pair production
    {\ee}, {\mm} and {\tautau}.}
  \label{fig:snlimit189}
\end{figure}

The exchange of scalar neutrinos can produce resonance
peaks at LEP energies. 
From an analysis of our measurements in the leptonic channels
upper limits on the coupling strength $ \lambda$
as a function of the scalar neutrino mass are determined.
The results for the {\EE}, {\mm} and {\tautau} final states are shown in
Figure~\ref{fig:snlimit189}.
In all  cases, large and previously unexplored areas in the 
$ (m_{\SNM}, \ \lambda_{121})$, $ (m_{\SNT}, \ \lambda_{131})$,   
$ (m_{\SNT}, \ \sqrt{\lambda_{131}\lambda_{232}})$ and
$ (m_{\SNM}, \ \sqrt{\lambda_{121}\lambda_{233}})$ planes are
excluded.

From the analysis of the hadronic cross section measurements,
upper limits on the Yukawa couplings $|\lambda'_{1jk}|$ $(j,k = 1,2,3)$
are derived
depending on the mass of exchanged scalar quarks. 
One single Yukawa coupling at a time is assumed to be much larger than the
others which are neglected.
Two cases are analysed:
 \begin{eqnarray*}
  m_{\tilde{U}} & \gg       & m_{\tilde{D}}         ~~~~\mathrm{with}~~
  \tilde{U}= \mathrm{\tilde{u}, \tilde{c}, \tilde{t}}~~\mathrm{and}~~
  \tilde{D}= \mathrm{\tilde{d}, \tilde{s}, \tilde{b}}  \\
  m_{\tilde{U}} & \muchless & m_{\tilde{D}} \: .
\end{eqnarray*}
Only the exchange of the much lighter scalar quark type 
is important. 
Due to quark  universality the limits on $|\lambda'_{1jk}|$  %  derived 
coincide  for each of the two cases of mass relation.

The R--parity breaking Yukawa couplings are mainly restricted by virtual
exchange of right--handed scalar down--type quarks in the $u$--channel
which couple in the same way as $S_0$ leptoquarks with  $|\gL|$.
Their amplitudes interfere with the equal--helicity amplitude (LL) of the 
Standard Model.

The amplitudes for left--handed scalar up--type quark exchange in the 
$t$--channel are the same as for $\tilde{S}_{1/2}$ leptoquark exchange
and interfere with the opposite--helicity amplitude (LR). The latter  
is suppressed in comparison to (LL).
The results on $|\lambda'_{1jk}|$  can be taken from Figure~\ref{fig:sca}
considering
  $S_0$(L) and $\tilde{S}_{1/2}$. 
Assuming scalar up--type and down--type quark masses to
be equal and both contributing to the hadronic cross section yields 
similar limits as for the case $m_{\tilde{U}} \gg m_{\tilde{D}}$. 

\section*{{\TeV} Strings}

Recently, it has been realized that in a string theory of quantum
gravity~\cite{Antoniadis,Peskin}
there are new phenomenological consequences.
For instance, massive string mode oscillations can lead to contact interactions,
which may have stronger effects than those
caused by the virtual exchange of gravitons.

The effects of {\TeV} scale strings on Bhabha scattering are computed~\cite{Peskin}
by multiplying the leading-order scattering amplitudes by a common
form factor, which depends on the string scale $ M_S$ and
the Mandelstam variables $s$ and $t$.
The Standard Model 
cross section for Bhabha scattering is
modified as follows:
\begin{equation}
 \frac{d \sigma}{d \COST} = \left(\frac{d \sigma}{d \COST}\right)_{SM} \left|\frac{\Gamma(1-\frac{s}{M_S^2})\Gamma(1-\frac{t}{M_S^2})}{\Gamma(1-\frac{s}{M_S^2}-\frac{t}{M_S^2})}\right|^2 ,
\end{equation}
where $\Gamma$ is the gamma function.

\begin{figure}[htbp]
  \begin{center}
     \includegraphics[width=0.999\textwidth]{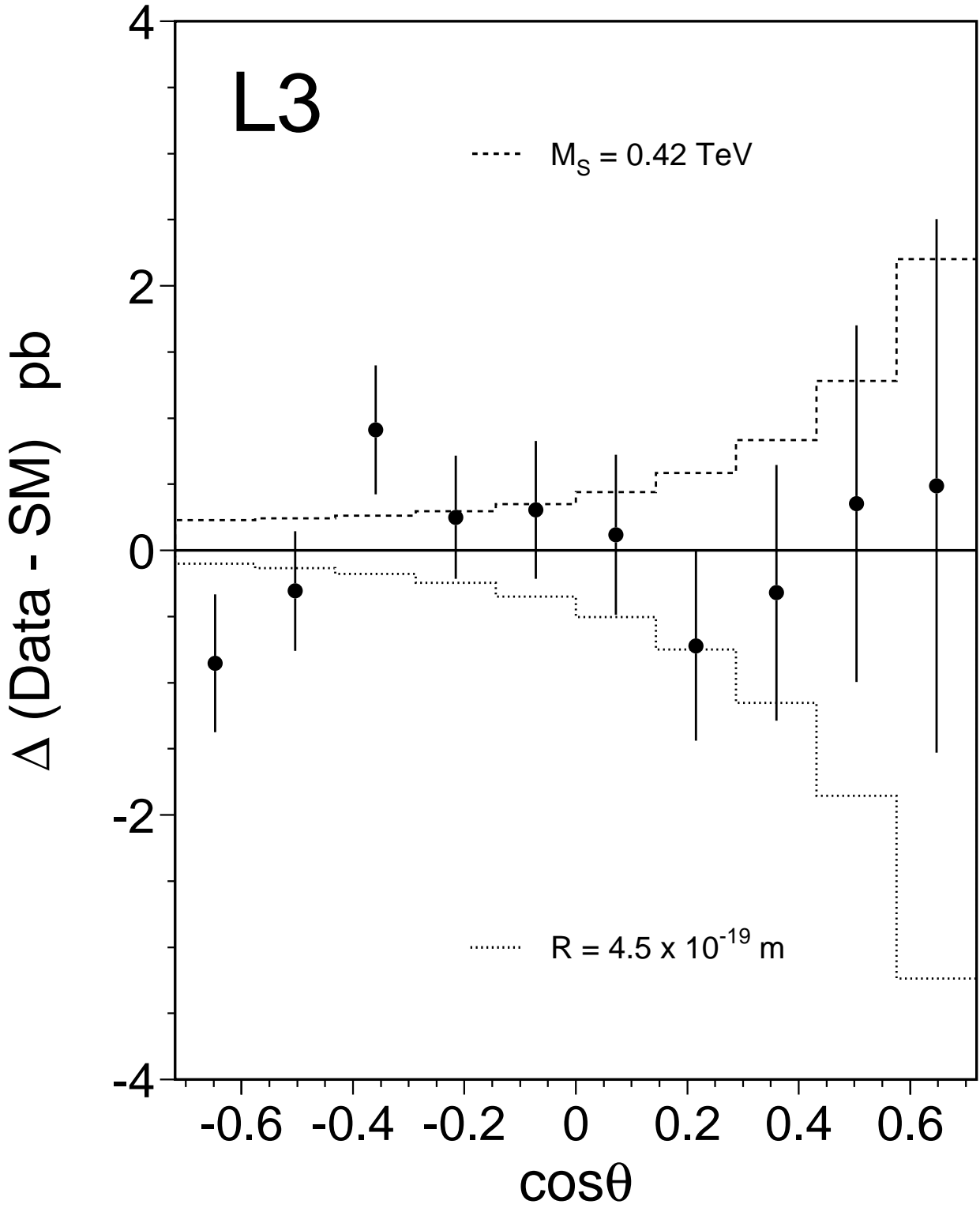}
  \end{center}
  \caption[]{
     Deviations of the measured differential cross section for Bhabha scattering
     at \mbox{$\sqrt{s}$ = 189 {\GeV}} from the Standard Model prediction.
     The effects expected in string models and for non--zero electron size
     are also shown.}
  \label{fig:string}
\end{figure}

The differential cross sections measured
at 183 and 189 {\GeV} are used to derive a lower limit on the
string scale  
$M_S$ of 0.49~{\TeV}.
The result of the analysis at 189 {\GeV} is depicted in Figure~\ref{fig:string}.

\section*{Form Factors and Fermion Sizes}

In the Standard Model the fermions and the gauge bosons are
considered to be pointlike.
If this is not the case, form factors or  anomalous
magnetic dipole moments of the fermions could be
observed~\cite{Zerwas:1995}.

The fermion-pair measurements
above the Z pole are analysed for such effects.
The Standard Model
cross sections for the reactions {\EEFF}
are modified as follows:
\begin{equation}
 \frac{d \sigma}{d q^2} = \left(\frac{d \sigma}{d q^2}\right)_{SM} F_e^2(q^2) F_f^2(q^2),
\end{equation}
where $ q^2$ is the Mandelstam variable $s$ or $t$ for
$s$-- or $t$--channel exchange, and
the form factors of the initial and final state fermions
are denoted as $ F_e$ and $ F_f$, respectively.
They are parametrized by a Dirac form factor:
\begin{equation}
 F(q^2) = 1 + \frac{1}{6} q^2 R^2,
\end{equation}
where $ R  $ is the radius of the fermion.

The upper limits on the fermion radii 
obtained from our data are shown in Table~\ref{tab:fradii}.
They are derived with the assumption $F_e = F_f$.
For the {\mm}, {\tautau} and {\QQ}
final states the limits given in Table~\ref{tab:fradii}
will increase by a factor of $\sqrt{2}$ under the
most conservative assumption that the electron is
pointlike ($F_e \equiv 1$).
The expected effects on the differential cross section
for the {\EE} final state are shown in Figure~~\ref{fig:string}.

The limits for quarks derived in this paper  are more stringent than
similar limits from high energy analyses of interactions
involving quarks and electrons by
the H1 collaboration~\cite{h1-00} in deep inelastic scattering,
and by the CDF collaboration~\cite{cdf-97} from study of the Drell-Yan process.

Limits on lepton radii have been extracted from 
the high precision low energy  measurements of
the magnetic dipole moment $ (g-2)$ of the electron and the
muon~\cite{Zerwas:1995,Brodsky}.
In the case where the deviations from the Standard Model of the magnetic
dipole moments of the leptons depend linearly on their mass,
the measurements of dipole moments can be interpreted as giving
much more stringent limits.
By contrast, in the case where the deviations depend quadratically on the
masses, our limit on the electron size is one order of magnitude lower,
and our limit on the muon size is similar to the limits
derived from $(g-2)$ measurements.

%%%%%%%%%%%%%%%%%%%%%%%%%%%%%%%%%%%%%%%%%%%%%%%%%%%%%%%%%%%%%%%%%%%%%%%%%%%%%%%%
%                            CONCLUSIONS
%%%%%%%%%%%%%%%%%%%%%%%%%%%%%%%%%%%%%%%%%%%%%%%%%%%%%%%%%%%%%%%%%%%%%%%%%%%%%%%%
\section*{Conclusions}

The measurements of fermion--pair cross sections and forward--backward asymmetries,
performed with the L3 detector at
centre--of--mass energies between 130~{\GeV} and 189~{\GeV},  are
used to search for effects of new physics phenomena.
No hint of manifestations of physics beyond the Standard Model is found.

The sensitivity of the searches, performed at energies above
the Z pole, has improved substantially compared to our previous
publications.
Limits on the energy scale $\Lambda$ of four--fermion contact
interactions in the range 3.8 -- 14.4~{\TeV} for leptons,
and in the range 2.8 -- 6.1~{\TeV} for quarks are obtained.
The effects of the exchange of leptoquarks or R--parity violating scalar 
quarks and scalar neutrinos are studied. 
In both cases, upper limits on the coupling constants, $|\gL|$ and $|\gR|$, 
or $|\lambda'|$ and $|\lambda|$
are determined as a function of the particle masses.
Lower limits on the mass of leptoquarks between 55 {\GeV}~and 560 {\GeV},
depending on the leptoquark type, are derived assuming
$\gLR = \sqrt{4 \pi \alpha}$.

In addition, new searches are performed for the effects of {\TeV}
strings, predicted in quantum gravity models, and a lower
limit on the string scale $ M_S$ of 0.49~{\TeV} is set.
From an analysis of form factors, upper limits on the
size of the different leptons and quarks in the
range \mbox{$(2.2 - 4.0) \, 10^{-19}$~m} are derived.

%
%%%%%%%%%%%%%%%%%%%%%%%%%%%%%%%%%%%%%%%%%%%%%%%%%%%%%%%%%%%%%%%%%%%%%%%%%%%%%%%
% Acknowledgements
%%%%%%%%%%%%%%%%%%%%%%%%%%%%%%%%%%%%%%%%%%%%%%%%%%%%%%%%%%%%%%%%%%%%%%%%%%%%%%%
%
\section*{Acknowledgements}

We are grateful to M.~Peskin for stimulating discussions.
We wish to 
express our gratitude to the CERN accelerator divisions for
the excellent performance of the LEP machine. 
We acknowledge the contributions of the engineers 
and technicians who have participated in the construction 
and maintenance of this experiment.  

%
%%%%%%%%%%%%%%%%%%%%%%%%%%%%%%%%%%%%%%%%%%%%%%%%%%%%%%%%%%%%%%%%%%%%%%%%%%%%%%%
% Bibliography
%%%%%%%%%%%%%%%%%%%%%%%%%%%%%%%%%%%%%%%%%%%%%%%%%%%%%%%%%%%%%%%%%%%%%%%%%%%%%%
%
% Style file to use with mcite.
% Use l3style with just cite.

\newpage

%
%%%%%%%%%%%%%%%%%%%%%%%%%%%%%%%%%%%%%%%%%%%%%%%%%%%%%%%%%%%%%%%%%%%%%%%%%%%%%%
% Author List
%%%%%%%%%%%%%%%%%%%%%%%%%%%%%%%%%%%%%%%%%%%%%%%%%%%%%%%%%%%%%%%%%%%%%%%%%%%%%%
%
%\newpage

\typeout{   }     
\typeout{Using author list for paper 208 -- ? }
\typeout{$Modified: Tue May  2 13:45:26 2000 by clare $}
\typeout{!!!!  This should only be used with document option a4p!!!!}
\typeout{   }
%
%
%
%  L A T E X  version!!
%
%
% Make sure that the Lep package has been used!
%\input{Lep.sty}%
%
%\ifx\LepCalled\undefined%
%\typeout{     }%
%\typeout{!!!!!!!!!!!!!!!!!!!!!!!!!!!!!!!!!!!!!!!!!!!!!!!!!!!!!!!!!!!}%
%\typeout{Yikes.  You haven't used the Lep package!}%
%\typeout{Please put \protect\usepackage\protect{Lep\protect} in your preamble,
%         followed by}%
%\typeout{\protect\Lep\protect{1\protect} or \protect\Lep\protect{2\protect}}%
%\typeout{     }%
%\typeout{For now you will get a Lep phase 2 authorlist (may not be right!).}%
%\typeout{!!!!!!!!!!!!!!!!!!!!!!!!!!!!!!!!!!!!!!!!!!!!!!!!!!!!!!!!!!!}%
%\typeout{     }%
%\Lep{2}\fi%

\newcount\tutecount  \tutecount=0
\def\tutenum#1{\global\advance\tutecount by 1 \xdef#1{\the\tutecount}}
\def\tute#1{$^{#1}$}
\tutenum\aachen            % 1
\tutenum\nikhef            % 2
\tutenum\mich              % 3
\tutenum\lapp              % 4
\tutenum\basel             % 5
\tutenum\lsu               % 6
\tutenum\beijing           % 7
\tutenum\berlin            % 8
\tutenum\bologna           % 9 
\tutenum\tata              % 10
\tutenum\ne                % 11
\tutenum\bucharest         % 12
\tutenum\budapest          % 13
\tutenum\mit               % 14 
\tutenum\debrecen          % 15
\tutenum\florence          % 16
\tutenum\cern              % 17 
\tutenum\wl                % 18 
\tutenum\geneva            % 19
\tutenum\hefei             % 20
\tutenum\seft              % 21
\tutenum\lausanne          % 22
\tutenum\lecce             % 23
\tutenum\lyon              % 24
\tutenum\madrid            % 25
\tutenum\milan             % 26
\tutenum\moscow            % 27
\tutenum\naples            % 27
\tutenum\cyprus            % 29
\tutenum\nymegen           % 30
\tutenum\caltech           % 31
\tutenum\perugia           % 32
\tutenum\cmu               % 33
\tutenum\prince            % 34
\tutenum\rome              % 35
\tutenum\peters            % 36
\tutenum\potenza           % 37
\tutenum\salerno           % 38
\tutenum\ucsd              % 39
\tutenum\santiago          % 40
\tutenum\sofia             % 41 
\tutenum\korea             % 42
\tutenum\alabama           % 43
\tutenum\utrecht           % 44
\tutenum\purdue            % 45
\tutenum\psinst            % 46
\tutenum\zeuthen           % 47
\tutenum\eth               % 48
\tutenum\hamburg           % 49
\tutenum\taiwan            % 50
\tutenum\tsinghua          % 51

{
\parskip=0pt
\noindent
{\bf The L3 Collaboration:}
\ifx\selectfont\undefined%  old style font selection
 \baselineskip=10.8pt
 \baselineskip\baselinestretch\baselineskip
 \normalbaselineskip\baselineskip
 \ixpt
\else%                      new style font selection
 \fontsize{9}{10.8pt}\selectfont
\fi
\medskip
\tolerance=10000
\hbadness=5000
\raggedright
\hsize=162truemm\hoffset=0mm
\def\r{\rlap,}
\noindent

M.Acciarri\r\tute\milan\
P.Achard\r\tute\geneva\ 
O.Adriani\r\tute{\florence}\ 
M.Aguilar-Benitez\r\tute\madrid\ 
J.Alcaraz\r\tute\madrid\ 
G.Alemanni\r\tute\lausanne\
J.Allaby\r\tute\cern\
A.Aloisio\r\tute\naples\ 
M.G.Alviggi\r\tute\naples\
G.Ambrosi\r\tute\geneva\
% ...
\newpage
%\rule{\textwidth}{0.4pt}
\begin{list}{A}{\itemsep=0pt plus 0pt minus 0pt\parsep=0pt plus 0pt minus 0pt
                \topsep=0pt plus 0pt minus 0pt}
\item[\aachen]
 I. Physikalisches Institut, RWTH, D-52056 Aachen, FRG$^{\S}$\\
 III. Physikalisches Institut, RWTH, D-52056 Aachen, FRG$^{\S}$
\item[\nikhef] National Institute for High Energy Physics, NIKHEF, 
     and University of Amsterdam, NL-1009 DB Amsterdam, The Netherlands
\item[\mich] University of Michigan, Ann Arbor, MI 48109, USA
\item[\lapp] Laboratoire d'Annecy-le-Vieux de Physique des Particules, 
     LAPP,IN2P3-CNRS, BP 110, F-74941 Annecy-le-Vieux CEDEX, France
\item[\basel] Institute of Physics, University of Basel, CH-4056 Basel,
     Switzerland
\item[\lsu] Louisiana State University, Baton Rouge, LA 70803, USA
\item[\beijing] Institute of High Energy Physics, IHEP, 
  100039 Beijing, China$^{\triangle}$ 
\item[\berlin] Humboldt University, D-10099 Berlin, FRG$^{\S}$
\item[\bologna] University of Bologna and INFN-Sezione di Bologna, 
     I-40126 Bologna, Italy
\item[\tata] Tata Institute of Fundamental Research, Bombay 400 005, India
\item[\ne] Northeastern University, Boston, MA 02115, USA
\item[\bucharest] Institute of Atomic Physics and University of Bucharest,
     R-76900 Bucharest, Romania
\item[\budapest] Central Research Institute for Physics of the 
     Hungarian Academy of Sciences, H-1525 Budapest 114, Hungary$^{\ddag}$
\item[\mit] Massachusetts Institute of Technology, Cambridge, MA 02139, USA
\item[\debrecen] KLTE-ATOMKI, H-4010 Debrecen, Hungary$^\P$
\item[\florence] INFN Sezione di Firenze and University of Florence, 
     I-50125 Florence, Italy
\item[\cern] European Laboratory for Particle Physics, CERN, 
     CH-1211 Geneva 23, Switzerland
\item[\wl] World Laboratory, FBLJA  Project, CH-1211 Geneva 23, Switzerland
\item[\geneva] University of Geneva, CH-1211 Geneva 4, Switzerland
\item[\hefei] Chinese University of Science and Technology, USTC,
      Hefei, Anhui 230 029, China$^{\triangle}$
\item[\seft] SEFT, Research Institute for High Energy Physics, P.O. Box 9,
      SF-00014 Helsinki, Finland
\item[\lausanne] University of Lausanne, CH-1015 Lausanne, Switzerland
\item[\lecce] INFN-Sezione di Lecce and Universit\'a Degli Studi di Lecce,
     I-73100 Lecce, Italy
\item[\lyon] Institut de Physique Nucl\'eaire de Lyon, 
     IN2P3-CNRS,Universit\'e Claude Bernard, 
     F-69622 Villeurbanne, France
\item[\madrid] Centro de Investigaciones Energ{\'e}ticas, 
     Medioambientales y Tecnolog{\'\i}cas, CIEMAT, E-28040 Madrid,
     Spain${\flat}$ 
\item[\milan] INFN-Sezione di Milano, I-20133 Milan, Italy
\item[\moscow] Institute of Theoretical and Experimental Physics, ITEP, 
     Moscow, Russia
\item[\naples] INFN-Sezione di Napoli and University of Naples, 
     I-80125 Naples, Italy
\item[\cyprus] Department of Natural Sciences, University of Cyprus,
     Nicosia, Cyprus
\item[\nymegen] University of Nijmegen and NIKHEF, 
     NL-6525 ED Nijmegen, The Netherlands
\item[\caltech] California Institute of Technology, Pasadena, CA 91125, USA
\item[\perugia] INFN-Sezione di Perugia and Universit\'a Degli 
     Studi di Perugia, I-06100 Perugia, Italy   
\item[\cmu] Carnegie Mellon University, Pittsburgh, PA 15213, USA
\item[\prince] Princeton University, Princeton, NJ 08544, USA
\item[\rome] INFN-Sezione di Roma and University of Rome, ``La Sapienza",
     I-00185 Rome, Italy
\item[\peters] Nuclear Physics Institute, St. Petersburg, Russia
\item[\potenza] INFN-Sezione di Napoli and University of Potenza, 
     I-85100 Potenza, Italy
\item[\salerno] University and INFN, Salerno, I-84100 Salerno, Italy
\item[\ucsd] University of California, San Diego, CA 92093, USA
\item[\santiago] Dept. de Fisica de Particulas Elementales, Univ. de Santiago,
     E-15706 Santiago de Compostela, Spain
\item[\sofia] Bulgarian Academy of Sciences, Central Lab.~of 
     Mechatronics and Instrumentation, BU-1113 Sofia, Bulgaria
\item[\korea]  Laboratory of High Energy Physics, 
     Kyungpook National University, 702-701 Taegu, Republic of Korea
\item[\alabama] University of Alabama, Tuscaloosa, AL 35486, USA
\item[\utrecht] Utrecht University and NIKHEF, NL-3584 CB Utrecht, 
     The Netherlands
\item[\purdue] Purdue University, West Lafayette, IN 47907, USA
\item[\psinst] Paul Scherrer Institut, PSI, CH-5232 Villigen, Switzerland
\item[\zeuthen] DESY, D-15738 Zeuthen, 
     FRG
\item[\eth] Eidgen\"ossische Technische Hochschule, ETH Z\"urich,
     CH-8093 Z\"urich, Switzerland
\item[\hamburg] University of Hamburg, D-22761 Hamburg, FRG
\item[\taiwan] National Central University, Chung-Li, Taiwan, China
\item[\tsinghua] Department of Physics, National Tsing Hua University,
      Taiwan, China
\item[\S]  Supported by the German Bundesministerium 
        f\"ur Bildung, Wissenschaft, Forschung und Technologie
\item[\ddag] Supported by the Hungarian OTKA fund under contract
numbers T019181, F023259 and T024011.
\item[\P] Also supported by the Hungarian OTKA fund under contract
  numbers T22238 and T026178.
\item[$\flat$] Supported also by the Comisi\'on Interministerial de Ciencia y 
        Tecnolog{\'\i}a.
\item[$\sharp$] Also supported by CONICET and Universidad Nacional de La Plata,
        CC 67, 1900 La Plata, Argentina.
\item[$\diamondsuit$] Also supported by Panjab University, Chandigarh-160014, 
        India.
\item[$\triangle$] Supported by the National Natural Science
  Foundation of China.
\end{list}
}
\vfill

%%% Local Variables: 
%%% mode: latex
%%% TeX-master: t
%%% End:

\newpage

%%%%%%%%%%%%%%%%%%%%%%%%%%%%%%%%%%%%%%%%%%%%%%%%%%%%%%%%%%%%%%%%%%%%%%%%%%%%%%%%
%                                TABLES
%%%%%%%%%%%%%%%%%%%%%%%%%%%%%%%%%%%%%%%%%%%%%%%%%%%%%%%%%%%%%%%%%%%%%%%%%%%%%%%%
\clearpage

\begin{table}
 \renewcommand{\arraystretch}{1.2}
  \begin{center}
    \begin{tabular}{|c|ccccccccc|}
\hline
~~Model~~&~~~LL~~&~~~RR~~&~~~LR~~&~~~RL~~&~~~VV~~
&~~~AA~~&~~~V0~~&~~~A0~~&LL$-$RR\\
  \hline \hline
$\eta_{\LL}$
   & $\pm$1& 0    &    0 &    0 &$\pm$1&$\pm$1&$\pm$1&  0  &$\pm$1 \\
$\eta_{\RR}$
   & 0     &$\pm$1&    0 &    0 &$\pm$1&$\pm$1&$\pm$1&  0  &$\mp$1  \\
$\eta_{\LR}$
   & 0     & 0    &$\pm$1&    0 &$\pm$1&$\mp$1& 0   &$\pm$1& 0   \\
$\eta_{\RL}$
   & 0     & 0    &    0 &$\pm$1&$\pm$1&$\mp$1& 0   &$\pm$1&  0 \\
\hline
    \end{tabular}
  \end{center}
  \caption{
    Models of contact interaction considered. The parameters $\eta_{ij}~
    (i,j=\mathrm{L,R})$ define to which helicity amplitudes
    the contact interactions contribute.}
  \label{tab:ci-models}
\end{table}
\begin{table} 
 \renewcommand{\arraystretch}{1.2}
  \begin{center}
    \begin{tabular}{|c||cc|cc|cc||cc|}
\hline
 Model     & \multicolumn{2}{|c|}{$\EE$}
           & \multicolumn{2}{ c|}{$\mu^+\mu^-$}
           & \multicolumn{2}{ c||}{$\tau^+\tau^-$}
           & \multicolumn{2}{|c|}{$\ll$} \\
   &~~~$\Lambda_-$~~~
         &~~~$\Lambda_+$~~~
               &~~~$\Lambda_-$~~~
                     &~~~$\Lambda_+$~~~
                           &~~~$\Lambda_-$~~~
                                  &~~~$\Lambda_+$~~~
                                         &~~~$\Lambda_-$~~~
                                                &~~~$\Lambda_+$~~~ \\
\hline  \hline
~~~~LL~~~~ & 4.9 & 4.3 & 3.8 &  8.5 & 4.7 & 5.4 & 5.2 & 9.0 \\
RR         & 4.9 & 4.3 & 3.6 &  8.1 & 4.4 & 5.1 & 5.1 & 8.7 \\
\hline                                                        
LR         & 5.8 & 5.1 & 2.0 &  6.5 & 1.8 & 3.7 & 6.4 & 6.3 \\
RL         & 5.8 & 5.1 & 2.0 &  6.5 & 1.8 & 3.7 & 6.4 & 6.3 \\
\hline                                                        
VV         &10.1 & 9.6 & 6.5 & 14.4 & 9.5 & 7.6 &10.3 &14.4 \\
AA         & 5.4 & 6.8 & 6.7 &  9.7 & 5.5 & 8.6 & 7.1 &12.4 \\
\hline                                                        
V0         & 6.8 & 6.3 & 5.4 & 11.7 & 6.6 & 7.3 & 7.3 &12.5 \\
A0         & 8.0 & 7.5 & 2.1 &  9.0 & 1.9 & 5.0 & 9.0 & 8.9 \\
LL$-$RR    & 3.0 & 3.0 & 3.5 &  4.4 & 2.9 & 3.3 & 3.8 & 4.6 \\
\hline
\end{tabular}
  \end{center}
   \caption{  
    The one--sided 95\% confidence level lower limits on the parameter $\Lambda$
    of contact interaction derived from fits to lepton--pair cross sections
    and asymmetries.
    The limits $\Lambda_{+}$ and $\Lambda_{-}$ given in {\TeV}  
    correspond to the upper and lower signs of 
    the parameters $\eta_{ij}$ in Table~\ref{tab:ci-models}.
    }
  \label{tab:ci-leptons}
\end{table}
\begin{table} 
 \renewcommand{\arraystretch}{1.2}
  \begin{center}
    \begin{tabular}{|c||cc||cc|cc||cc|}
\hline
 Model     &\multicolumn{2}{|c||}{$\QQ $}
           &\multicolumn{2}{ c|}{$\UU$}
           &\multicolumn{2}{ c||}{$\DD$}     
           &\multicolumn{2}{|c|}{$\FF$ }\\
           &~~~$\Lambda_-$~~~
                 &~~~$\Lambda_+$~~~
                       &~~~$\Lambda_-$~~~
                             &~~~$\Lambda_+$~~~
                                   &~~~ $\Lambda_-$~~~
                                          &~~~$\Lambda_+$~~~
                                               &~~~ $\Lambda_-$~~~
                                                     &~~~$\Lambda_+$~~~  \\
\hline \hline  
~~~~LL~~~~ & 2.8 & 4.2 & 4.1 & 7.0 & 6.5 & 3.9 & 5.5 & 8.3 \\ 
    RR     & 3.8 & 3.1 & 3.6 & 1.5 & 1.8 & 2.9 & 4.9 & 9.0 \\ 
\hline                                                       
    LR     & 3.5 & 3.3 & 2.7 & 1.9 & 2.1 & 2.5 & 5.9 & 6.1 \\ 
    RL     & 4.6 & 2.5 & 2.4 & 2.2 & 2.8 & 1.8 & 6.0 & 7.9 \\ 
\hline                                                       
    VV     & 5.5 & 4.2 & 5.8 & 9.8 & 2.2 & 4.6 & 9.8 &15.0 \\ 
    AA     & 3.8 & 6.1 & 5.0 & 7.4 & 7.5 & 5.1 & 7.6 &11.3 \\ 
\hline                                                       
    V0     & 3.7 & 4.4 & 5.2 & 9.2 & 7.7 & 4.7 & 7.2 &12.1 \\ 
    A0     & 5.2 & 3.1 & 3.2 & 2.3 & 2.9 & 2.5 & 8.3 &10.0 \\ 
LL$-$RR    & 3.2 & 5.1 & 3.2 & 2.3 & 2.1 & 3.6 & 3.5 & 5.5 \\ 
\hline
    \end{tabular}
  \end{center}
  \caption{
    The one--sided 95\% confidence level lower limits on the parameter $\Lambda$
    of contact interaction derived from fits to hadron cross sections,
    and for all fermions combined.
    The limits $\Lambda_{+}$ and $\Lambda_{-}$ given in {\TeV}
    correspond to the upper and lower signs of 
    the parameters $\eta_{ij}$ in Table~\ref{tab:ci-models}.
    }
  \label{tab:ci-hadrons}
\end{table}
\begin{table}
 \renewcommand{\arraystretch}{1.2}
  \begin{center}
   \begin{tabular}{|l|r||l|r|}
    \hline 
        LQ type  & $m_{\mathrm{LQ}}$[\GeV]
      & LQ type  & $m_{\mathrm{LQ}}$[\GeV] \\
    \hline
    $S_0(\mathrm{L})\rightarrow \mathrm{e u}$                      &  390  &
    $V_{1/2}(\mathrm{L})\rightarrow \mathrm{e d}$                  &  190  \\
    $S_0(\mathrm{R})\rightarrow \mathrm{e u}$                      &  300  &
    $V_{1/2}(\mathrm{R})\rightarrow \mathrm{e u,~ e d}$            &  170  \\
    $\tilde{S}_0(\mathrm{R})\rightarrow \mathrm{e d}$              &  80   &
    $\tilde{V}_{1/2} (\mathrm{L}) \rightarrow \mathrm{e u}$        &  140 \\
    $S_1(\mathrm{L})\rightarrow \mathrm{e u,~ ed}$                 &  200  &
    $V_0(\mathrm{L})\rightarrow \mathrm{e \bar{d}}$                &  560  \\
    $S_{1/2}(\mathrm{L})\rightarrow \mathrm{e \bar{u}}$            &  55   &
    $V_0(\mathrm{R})\rightarrow \mathrm{e \bar{d}}$                &  130  \\
    $S_{1/2}(\mathrm{R})\rightarrow \mathrm{e \bar{u},~e \bar{d}}$ &  110  &
    $\tilde{V}_0(\mathrm{R})\rightarrow \mathrm{e \bar{u}}$        &  280 \\
    $\tilde{S}_{1/2}(\mathrm{L})\rightarrow \mathrm{e \bar{d}}$    & --    &
    $V_1(\mathrm{L})\rightarrow \mathrm{e \bar{u},~e \bar{d}} $    & 380 \\
    \hline
   \end{tabular}
  \end{center}
  \caption{ 
   Lower limits on the mass of leptoquarks 
   at 95\% confidence level derived from hadronic
   cross section measurements assuming
   $g=\sqrt{4 \pi \alpha}$.}
  \label{tab:lqmass}    
\end{table}
\begin{table}
 \renewcommand{\arraystretch}{1.2}
  \begin{center}
    \begin{tabular}{|c|c|}
\hline
~~Channel~~&~~~$R$  [m]~~\\
\hline 
\ee\       & $\rm 3.1\cdot 10^{-19}$ \\
\mm\       & $\rm 2.4\cdot 10^{-19}$ \\
\tautau\   & $\rm 4.0\cdot 10^{-19}$ \\
\hline
$\ll$      & $\rm 2.2\cdot 10^{-19}$ \\
\hline
\QQ\       & $\rm 3.0\cdot 10^{-19}$ \\
\hline
    \end{tabular}
  \end{center}
  \caption{
    Upper limits on the fermion radii at 95~\% confidence level
    for electrons, muons, taus, all leptons combined, and for 
    quarks.}
  \label{tab:fradii}
\end{table}

%%%%%%%%%%%%%%%%%%%%%%%%%%%%%%%%%%%%%%%%%%%%%%%%%%%%%%%%%%%%%%%%%%%%%%%%%%%%%%%%
%                               FIGURES
%%%%%%%%%%%%%%%%%%%%%%%%%%%%%%%%%%%%%%%%%%%%%%%%%%%%%%%%%%%%%%%%%%%%%%%%%%%%%%%%
%\clearpage
\newpage

%%%%%%%%%%%%%%%%%%%%%%%%%%%%%%%%%%%%%%%%%%%%%%%%%%%%%%%%%%%%%%%%%%%%%%%%%%%%%%%%
%                              DOCUMENT END
%%%%%%%%%%%%%%%%%%%%%%%%%%%%%%%%%%%%%%%%%%%%%%%%%%%%%%%%%%%%%%%%%%%%%%%%%%%%%%%%
\end{document}